\documentclass[prl,nofootinbib,superscriptaddress,twocolumn,showpacs]{revtex4-1}
 
\usepackage{bm}
\usepackage{amsfonts}
\usepackage{amssymb}
\usepackage{amsmath}
\usepackage{amsthm}
\usepackage{hyperref}
\usepackage{times}

\usepackage{subfigure}
\usepackage[pdftex]{color,graphicx}

\newcommand{\ket}[1]{\ensuremath{\left| #1 \right\rangle}}
\newcommand{\bra}[1]{\ensuremath{\left\langle #1 \right|}}

\newcommand{\nufh}{{$\nu=5/2$}}

\newcommand{\avg}[1]{\left< #1 \right>}

\let\oldmarginpar\marginpar
\renewcommand\marginpar[1]{\-\oldmarginpar[\raggedleft\tiny\color{red} #1]%
{\raggedright\tiny #1}}

\graphicspath{{Figures/}}

\begin{document}

\title{Detecting Non-Abelian Anyons by Charging Spectroscopy}
	
\author{G. Ben-Shach}
\affiliation{Department of Physics, Harvard University, Cambridge, MA 02138}
\author{C. R. Laumann}
\affiliation{Department of Physics, Harvard University, Cambridge, MA 02138}

\author{I. Neder}
\affiliation{Raymond and Beverly Sackler School of Physics and Astronomy, Tel-Aviv University, Tel Aviv, 69978, Israel}
\author{A. Yacoby}
\affiliation{Department of Physics, Harvard University, Cambridge, MA 02138}
\author{B. I. Halperin}
\affiliation{Department of Physics, Harvard University, Cambridge, MA 02138}

\date{\today}

\begin{abstract}
	Observation of non-Abelian statistics for the $e/4$ quasiparticles in the $\nu=\frac{5}{2}$ fractional quantum Hall state remains an outstanding experimental problem. The  non-Abelian statistics are linked to the presence of additional low energy states in a system with localised quasiparticles, and hence an additional low-temperature entropy. Recent experiments, which detect changes in the number of quasiparticles trapped in a local potential well as a function of an applied gate voltage, $V_G$, provide a possibility for measuring this entropy, if carried out over a suitable range of temperatures, $T$. We present a microscopic model for quasiparticles in a potential well and study the effects of non-Abelian statistics on the charge stability diagram in the $V_G-T$ plane, including broadening at finite temperature. We predict a measurable slope for the first quasiparticle charging line, and an \textit{even-odd} effect in the diagram, which is a signature of non-Abelian statistics. 
\end{abstract}

\pacs{73.43.Cd, 05.30.Pr, 71.10.Pm}

\maketitle


%
The unambiguous observation of particles obeying non-Abelian statistics remains an outstanding experimental challenge in condensed matter physics.
The Moore-Read fractional quantum Hall state (FQH)~\cite{MooreRead}, believed to be realized at filling fraction \nufh{}, is one of the most promising candidate phases to exhibit such quasiparticles (QPs)~\cite{NayakRMP}. 
The Moore-Read state is predicted to support QPs of charge $\pm e/4$; for $N$ such QPs, localized and well separated from each other, there should be a nearly degenerate set of ground states, with multiplicity $2^{N/2-1}$.
For temperature $T$ larger than the splitting of these ground states, but smaller than the gap to higher excited states, this degeneracy contributes an effective entropy to the system, the \textit{non-Abelian entropy}.

Non-Abelian statistics predicts that pairs of QPs can interact to form two distinct states, or \textit{fusion channels}, $f$, commonly denoted as $f=1,~\psi$.
In a finite system, the two states have different energies, and the ground state is unique; for $T$ below the splitting between the two, the non-Abelian entropy is lost.

\begin{figure}[htbp]
	\centering
		\includegraphics{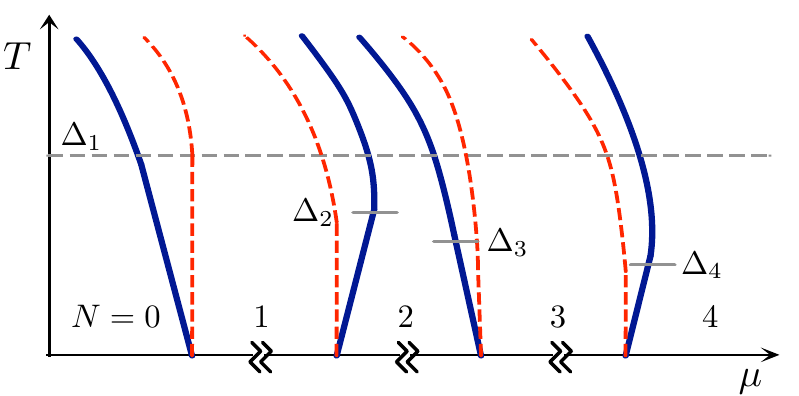}
	\caption{Cartoon charge stability diagram, showing only peak centres (no broadening). Vertical axis is temperature $T$; horizontal axis $\mu$ is the chemical potential for charged QPs, controlled in experiment by a gate potential. Red dashed lines are for Abelian particles. Blue solid lines correspond to non-Abelian QPs in a tightly confining well. $\Delta_N$ is the gap to excited states for $N$ particles, and sets scale for other entropic effects. Notice even-odd effect for non-Abelian anyons. }
	\label{fig:Charge_Stability}
\end{figure}
There are several recent theoretical proposals for techniques to observe this entropy through bulk measurement of thermodynamic and transport properties~\cite{PhysRevB.79.115317,PhysRevLett.102.176807,PhysRevLett.105.086801,Laumann}.
Recent measurements in this direction of thermoelectric response at \nufh{} are encouraging~\cite{Eisenstein2012}. 
These theoretical proposals assume that all QPs are well separated, such that degeneracy-lifting interactions are weak or non-existent.
However, recent local electronic charge-sensing measurements, using a single-electron transistor (SET)~\cite{Venkatachalam2011}, suggest that QPs tend to trap in local potential wells due to electrostatic disorder, which may be tightly confining and contain more than one QP.
Confined QPs split their degeneracy through two means: Majorana exchange~\cite{ChengdasSarma,Baraban}, present even for stationary QPs, and as we show here, an orbital splitting from interchange of the charged QPs, which can dominate in special cases.

In this letter, we study the charging spectra of local quasiparticle traps.
Such traps may be induced by disorder or defined by gates.
Their spectra reflect the QP statistics, just as electronic dot spectra reflect the spin and fermionic statistics of electrons.
We show that low-frequency SET charge-sensing measurements, which provide only thermally-averaged information regarding the dot spectra, are sufficient for extracting non-Abelian signatures.
At low but experimentally accessible $T$, we predict a robust temperature evolution of the $N=0-1$ transition, and an even-odd effect in the evolution of the charging spectrum for several non-Abelian anyons. 
This effect should be visible for $T$ below the relevant gaps to excited states for $N$ particles, which we calculate for $N=1,2$.

%
The experiments of~\cite{Venkatachalam2011} measure the change in potential at the SET induced by a change $\delta V_G$ in the potential applied to a backgate on the sample. 
If there is a single disorder-induced well close to the SET, the measured signal is inversely proportional to the compressibility of the well, $\kappa=\frac{\partial \avg{N}}{\partial \mu}$, where $\mu$ is the QP chemical potential in the vicinity of the well.
For an isolated well, the relation between $\delta V_G$ and the change in $\mu$ should be linear, but the constant of proportionality is geometry-dependent, as screening depends on the local environment as well as the distance to the gate~\cite{Supplement}.
If there are several wells nearby, their signals are weighted according to the strength of their coupling to the SET; in this case, Coulomb interactions between wells need also be taken into account.

At $T=0$, the compressibility has a $\delta$-function peak at a crossing of energy levels between $N$ and $N+1$ QPs in a well.
At finite $T$, the peak broadens and may shift as a function of $\mu$ due to entropy effects.
The simplest case to consider is an isolated well at the transition from $N=0$ to $N=1$, or slightly more involved, from one to two.
At higher occupation numbers, we give qualitative arguments for the stability diagram.
We examine both circular and elliptical traps, and account for temperature effects including broadening and excited states.
In our model, $e/4$ QPs are represented as interacting charged particles in a magnetic field, confined to the lowest Landau level (LLL), with non-Abelian statistics.
The interaction is Coulomb, supplemented by an interaction $V_X(r)$ due to the exchange of Majorana fermions.
%

%
%
\paragraph{Qualitative Picture}
We begin with the charging diagram for Abelian particles in a well, to contrast it with the non-Abelian case.
For simplicity of presentation, we consider varying only the QP chemical potential, although as discussed below, local gating will be required to access the full charging spectrum.  
The well sits in a larger quantum Hall state containing other distant wells, which provide a reservoir for QPs.
At $T=0$, as a function of chemical potential, a series of peaks in the compressibility appear, corresponding to individual charging events in the well. 
The spacing of these peaks defines the charging energy, $U(N)$. 
As $T$ increases, the peak centres evolve vertically in the charging diagram (red dashed lines in Fig~\ref{fig:Charge_Stability}), until $T$ reaches the minimum excitation energy, $\Delta_N$, set by the excited states within the well.
Above this energy, the curve deviates from a straight line due to entropic effects.
The peaks broaden linearly with $T$ for both Abelian and non-Abelian QPs.

When several non-Abelian QPs occupy a tightly confining well, they uniquely fuse at low energies.
This produces a distinct experimental signature - the \textit{even-odd} effect. 
As highlighted in~\cite{PhysRevLett.102.176807}, the density dependence of the zero-temperature entropy produces a distinct signature in the inverse compressibility of bulk samples at low $T$.
In local traps with discrete QP number, the difference in zero-temperature entropy $\Delta S$ between adjacent number states produces a related low-$T$ signature in the charge stability diagram: the slope of the charge transition line in the $\mu-T$ plane is $-1/\Delta S$. 
The first QP placed in the well contributes $S_{NA} = \ln 2/2$ to the non-Abelian entropy ($k_B = 1$), or equivalently adds a $\sqrt{2}$ degeneracy; thus,  the $N=0-1$ transition line has slope $-2/\ln 2$ in the $\mu-T$ plane as $T\rightarrow 0$.
A second QP fuses uniquely with the first QP into the $1$ or $\psi$ channel and the non-Abelian entropy is extinguished, $\Delta S = -\ln{2}/2$. 
Thus, as $T$ increases from zero, the $N=1$ state becomes entropically more favourable than the $N=2$ state, and the transition line has slope $+2/\ln 2$ (blue solid lines in Fig~\ref{fig:Charge_Stability}). 
This even-odd effect persists as the well charges: odd numbers of particles fuse into the non-Abelian $\sigma$-channel, while even numbers uniquely fuse into either the Abelian $1$ or $\psi$ channels, as long as $T$ remains below the splitting between these two channels.
If the splitting between channels is smaller than $\Delta_N$, there exists an intermediate regime, in which the degeneracy is preserved and the non-Abelian entropy increases by $\ln 2/2$ with every additional particle, and all lines have parallel negative slopes.
A similar effect for electrons due to spin degeneracy was predicted and seen in quantum dots at $B=0$~\cite{Bennett1,Bennett2}.

For wells far apart compared to the magnetic length, the rate, $\delta_{\text{well}}/\hbar$ of Majorana exchange between them falls off exponentially in their separation.
We therefore consider charging lines for temperatures $T \gg \delta_{\text{well}}$, assumed zero for an isolated well.
In this limit there is no fusion-channel splitting between wells, and each independently exhibits the even-odd effect. 
However, the charging spectra are not completely independent due to capacitive coupling. 
In an experiment sensitive to multiple disorder-induced wells, the charging spectra of the wells appear overlaid with unknown offsets making the even-odd effect more difficult to observe, without first associating the various peaks to their respective wells. 
Experiments~\cite{Venkatachalam2011,Ilani1,Jens1} suggest that determining such associations is possible.
%

%
%
\paragraph{Equilibration} 
Although QPs are locally trapped, the equilibrium model we present requires that the system explores the degenerate ground state manifold faster than the measurement time $t_{\text{exp}}$ of the charge-sensing experiments.
This time is determined by the rate of change of the gate voltage, typically $t_{\text{exp}} \sim 0.1\text{s}$.
We estimate the equilibration time due to thermal excitations as $t_{\text{T}} \sim 10^{-4}\text{s} \ll t_{\text{exp}}$, meeting the requirement~\cite{Supplement}.
Moreover, the observed changes in the charge state of the studied well during experiments~\cite{Venkatachalam2011} imply that QPs hop freely between wells on the time scale $t_{\text{exp}}$.
Assuming the hopping processes have a stochastic component, they will naturally lead to braiding of QPs from different wells.

\paragraph{Quantitative Picture}
Returning to a single well in a large bath, we present a model for calculating the charging diagram. 
The partition function is
\begin{align}
	\label{eq:partition}
	Z &= \sum_N g(N)e^{-\beta (F(N,T)-\mu N)},
\end{align}
where $\beta=1/T$, the internal free energy of $N$-particle states in the well is $F(N,T)$, and 
\begin{align}
	g(N) &= \left\{ \begin{array}{ll}   
		\sqrt{2} & N \textrm{ odd} \\
		1 & N \textrm{ even}
		   \end{array} \right.
\end{align}
captures the non-Abelian degeneracy associated with net fusion within the well.
In a well where QPs are close, such that all other fusion-degeneracies are split by energies larger than $T$, we take $F(N,T) \approx F(N,0)$ for $T\ll\Delta_N$, the gap to excitations. 
In principle, however, for a wide well where electron-electron interactions localise the QPs further apart, an intermediate regime can exist in which the topological degeneracies are not significantly split and $F(N,T) \approx F(N,0) - T \lfloor\frac{N}{2}\rfloor \ln{2}$, where $\lfloor \cdot \rfloor$ denotes the integer part, for temperatures up to the gap $\Delta_N$. 

The compressibility follows from the partition function. To leading order near the $N-1$ to $N$ charge transition at the critical chemical potential, $\mu_0^{N} \equiv F(N,0)-F(N-1,0)$,
\begin{equation}
	\kappa = \beta \frac{\frac{g(N)}{g(N-1)}e^{\beta (\delta \mu-\Delta F)}}{\left( 1+\frac{g (N)}{g(N-1)} e^{\beta (\delta \mu -\Delta F)} \right)^2},
\end{equation}
where $\delta \mu = \mu - \mu_0^{N}$ and $\Delta F = F(N,T) - F(N-1,T) - \mu_0^{N}$.
We differentiate with respect to $\delta \mu$ to find the centre of the peak: 
$\delta \mu_{\text{max}} = T \ln \left(g(N-1)/g(N)\right) + \Delta F $. 
For a tightly confining circular well at low $T$, for which $\Delta F = 0$, this gives $\delta \mu_{\text{max}}= \pm (T/2)\ln{2}$, which confirms that the charging line slopes alternate sign as a function of the parity of $N$.
In the intermediate regime, the slope is negative for all $N$. 
The peak height decreases with $T$ as $\kappa_{\text{max}} \sim 1 / 4 T$, while the full-width-half-max (FWHM) increases with $T$ due to number fluctuations as $\text{FWHM} \sim 2T\ln(3+2\sqrt{2})$, roughly ten times as fast as the shift in position. 
Nevertheless, tracing the peak should be possible if measurements are sufficiently accurate.
In the experimental regime of interest, the charging energy $U(N)=\mu_0^{N}-\mu_0^{N-1} \gg T$, so the peaks remain distinguishable.
The key input to the above statistical model is the microcanonical low-energy spectra of fixed numbers of QPs in a well, which we now calculate.

\paragraph{One Particle}
For a particle in an elliptical harmonic well,
\begin{align}
	V_{\text{trap}}&=\frac{1}{2}k\left(x^2 + \alpha y^2\right),
\end{align} 
where $k$ is the spring constant, and $\alpha$ controls the eccentricity ($\alpha=1$ defines a circular trap), the level-spacing is $\Delta_1 = k\sqrt{\alpha}{l_B^*}^2$, where $l_B^*=\sqrt{{\hbar}/{e^* B}}$ is the effective magnetic length for QPs in a magnetic field $B$. 
At finite $T$, this produces an internal free energy, 
\begin{align}
	F(1,T) = T \ln (1-e^{-\Delta_1/T}).
\end{align}
This free energy decreases weakly with $T$ for $T < \Delta_1$, only significantly correcting the linear charging curve for $T\gg\Delta_1$, as shown in figure~\ref{fig:Charge_Stability}.

\paragraph{Two Particles}
As the fusion channel, $f$, of two orbiting non-Abelian anyons is conserved, the orbital dynamics may be treated separately in each $f$-sector. 
This reduces to the dynamics of Abelian anyons whose statistical angle $\theta$ depends on the fusion sector.
For Ising anyons, $\theta_{1}=0$ and $\theta_{\psi}=\pi/2$~\cite{ParsaThesis}. 
To model two such anyons in a well, each with charge $e^*=e/4$, we write the Hamiltonian for a pair of bosons with a statistical gauge field:
\begin{align}
	H=& \frac{1}{2m} \sum_{i=1}^2 \left(\vec{p}_i-\hbar \vec{a}^f_i-e^*\vec{A}_i\right)^2 + V_{\text{trap}}(\vec{r}_i)+ \\ \nonumber
	&+V_{I}(\vec{r}_1-\vec{r}_2)+ V^f_{X}(\vec{r}_1-\vec{r}_2).
\end{align}
The first term contains the electromagnetic vector potential $\vec{A}_i$, corresponding to a uniform external $B$-field, as well as a statistical gauge field $\left(a^f_x,a^f_y\right)=\frac{\theta_f}{\pi r^2}\left(y,-x\right)$, which binds a flux tube of strength $\theta_f$ to each quasiparticle, and $m$ is the effective QP mass.
We project into the LLL, taking $m \rightarrow 0$. 
The coordinates in $\vec{a}_i$ are relative to the other particle. 
We assume that the QPs interact via a Coulomb interaction, $V_{I}=\frac{{e^*}^2}{4\pi \epsilon r}$, where $\epsilon \equiv \epsilon_r\epsilon_0$ is the electric permittivity of the material. 
This approximation is valid assuming that QPs do not come within $l_B^*$ of each other.
$V_X$ is the direct energy splitting of the fusion channels due to virtual exchange of Majorana fermions.
It is related to the fusion channel splitting discussed in~\cite{ChengdasSarma,Baraban}, and should consist of an exponential decay and oscillations, each on the order of several $l_B^*$.
For circular wells, the behaviour of Abelian anyons has been treated previously~\cite{SivanLevitANYONS, wen2004quantum, JohnsonCanright,SternRosenowIlanHalperin}. 
We summarise key results, and include corrections due to eccentricity.

In the symmetric gauge for harmonic traps, the centre-of-mass (CM) and relative (REL) coordinates decouple. 
In the CM coordinate, the statistical gauge field $\vec{a}^f$ falls out, leaving a single particle projected into the LLL in a harmonic well. 
For the REL coordinate, the particle is confined to a half-plane with the origin removed, and $\vec{a}^f$ remains~\cite{wen2004quantum}. 
We change the gauge, so that $\vec{a}^f=0$, giving a twisted boundary condition, $\psi_{REL}(r,\pi)=e^{i\theta_f}\psi_{REL}(r,0)$.
The potential landscape in the half-plane is defined by strong Coulomb repulsion near the origin together with the harmonic trap, $V_{\text{trap}}+V_{I}$, for $V_X=0$. 
The twisted periodic boundary conditions allow only angular momenta $\ell=2n+\theta/\pi$, for $n$ integer. 
The REL-coordinate wave-functions in the LLL have a basis given by $\ket{\ell}$,
\begin{equation}
	\left\langle z | \ell \right\rangle = N_{\ell}^{-\frac{1}{2}}z^{\ell}e^{-|z|^2/4(2{l_B^*}^2)},
\end{equation}
where $z=x+iy$ and $N_{\ell}$ is a normalisation constant on the half-plane. 
In this basis, we can diagonalise to find the two-particle spectrum. 
The potential has diagonal terms, as well as an off-diagonal term only when circular symmetry is broken~\cite{Supplement}. 

\paragraph{Circular Well}
We assume $V_{X}=0$ initially, and note that the CM coordinate behaves just like the single particle case with $\Delta_{\text{CM}} = \Delta_1$.
The lowest energy gap $\Delta^f_{\text{R}}$ in the relative coordinate within a fusion channel $f$ can be found by taking differences between adjacent $\ell$-states near the minimum, obtained by diagonalising the Hamiltonian.
We define the parameter $r_0=({{e^*}^2}/{2\pi \epsilon k} )^{1/3}$, the radial position of the minimum of the potential.
This splitting $\Delta^f_{\text{R}}$ oscillates with $r_0$ at fixed magnetic field with an amplitude that decays in the large-well limit, $r_0 \gg l_B^*$, as
\begin{align}
	\Delta^f_{\text{R}} & \lesssim 12 \Delta_1 \frac{r_0^2}{{l_B^*}^2} = 24 \frac{{e^*}^2}{4\pi \epsilon} \frac{{l_B^*}^4}{r_0^5},
\end{align}
The other relevant gap for the relative coordinate is the energy difference $E_{1\psi} = |E_0^{1}-E_0^{\psi}|$ between lowest energy states in the $1$ and $\psi$ channels.
With $V_{X}=0$, the splitting between fusion channels is an interchange effect, which follows from the allowed angular momenta in each channel; in particular, $E_{1\psi}$ behaves similarly to $\Delta^f_{\text{R}}$ with a maximum oscillation bounded by the power law $\frac{9}{2} \frac{{e^*}^2}{4\pi \epsilon} \frac{{l_B^*}^4}{r_0^5}$, which is approximately $20\%$ of the amplitude of $\Delta^f_{\text{R}}$.
For $T < E_{1\psi}$ and $\Delta^f_{\text{R}}$, the slope of the $1$-$2$ transition in the $\mu-T$ plane is positive, exhibiting the even-odd effect. 
Clearly, intra-channel entropy washes out the effect for $T > \Delta^f_{\text{R}}$. 
As $r_0$ varies, $E_{1\psi}$ will oscillate in sign, and can be arbitrarily small if $r_0$ is close to a zero-crossing. 
If $E_{1\psi} < T < \Delta^f_{\text{R}}$, the $0$-$1$ and $1$-$2$ charging lines are parallel with negative slope $-2/\ln 2$~\cite{Supplement}.

Non-Abelian QPs at finite separation can exchange Majorana fermions, leading to an additional fusion channel splitting.
Unlike the orbital contribution, this splitting occurs even when QPs are localised. 
Using a variational method to calculate this energy splitting for particles on a sphere, it was found to decay exponentially on the order of several magnetic lengths, up to a numerical pre-factor of $\mathcal{O}(1)$~\cite{Baraban}.
$V_{X}$ in the Hamiltonian accounts for a splitting of this form.
We do not calculate $V_X$ explicitly, but note that while it dominates the shift between fusion channel spectra in tightly confining wells, it oscillates and decays exponentially as the well widens and particle separation increases.
In general, $V_{X}$ increases $E_{1\psi}$, promoting the even-odd effect over a larger $T$-range, and making a regime of parallel charging lines less likely.
%

\paragraph{Anisotropic Well}
For anisotropic wells, again taking $V_{X}=0$ initially, consider the relative coordinate for two QPs.  
Starting from the circular well where QP orbits encircle the origin in the half-plane, as the eccentricity $\alpha$ increases, the effective potential acquires a minimum on the $x$-axis, at $x=r_0$, and a saddle point on the $y$ axis at $y= r_0/\alpha^{1/3}$.
For any given $\alpha>1$, the wavefunction becomes effectively localised near the potential minimum for $(r_0/l_B^*)^2 > \frac{2}{\sqrt{3}}\frac{\sqrt{\alpha-1}}{(\alpha^{1/3}-1)} \equiv \lambda(\alpha)$.
This is when the lowest-energy state near the minimum has energy lower than the saddle point potential.
As $\alpha \rightarrow 1$, $\lambda(\alpha)$ diverges as $(\alpha-1)^{-1/2}$, confirming that for a circular well, QPs are not localised.
Eccentricity breaks any accidental degeneracies which arise in the circular potential near $r_0$, and modifies the spectrum of the well.
For low eccentricities, the degeneracy breaking can increase or decrease the orbital splitting.
For large enough $\alpha$, the QPs are trapped at opposite ends of the well, and no longer orbit each other, except for quantum tunneling across the saddle point.
In a saddle point tunneling model, the orbital exchange rate, $R$, in the large well limit is Gaussian in the well-size, $R\approx  k {l_B^*}^2 \exp [ -\alpha^{-1/2} \lambda(\alpha)^{-1}c(\alpha)\left(r_0/l_B^*\right)^2 ]$, where $c(\alpha)$ depends weakly on $\alpha$ and goes to a constant of order unity as $\alpha \rightarrow 1$~\cite{Supplement}. 
This expression may be obtained by estimating the potential as Harmonic near the minimum, and using a WKB type calculation of the tunneling of a particle near a quadratic saddle point in the LLL, as in~\cite{FertigHalperin}.
Increasing $\alpha$ also has the effect of raising the energy of the ground state, by increasing the harmonic frequency of the trap. 

For anisotropic wells with $V_X \neq 0$, the exchange effect naturally dominates the splitting at large $r_0$, since the exchange of neutral Majorana fermions decays exponentially while the interchange of localised charged particles in a magnetic field decays as a Gaussian. 
We recover the even-odd effect for $T$ below this splitting, regardless of QP localisation.

%
\paragraph{Energy Estimates}
A simple model producing a charge trap is provided by considering a point-like gate, a distance $d$ above the 2DEG. 
A charge $+|e|$ on this gate produces an effective circular harmonic trap in the plane with spring constant $k=\frac{|e e^*|}{4\pi \epsilon d^3}$.
Using $\epsilon_r=13$ for GaAs/AlGaAs quantum wells, $B = 3.5 \text{T}$ and $d = 100 \text{nm}$, we find $r_0= 63\text{nm}$.
The charging energy is $1.6\text{K}$, and the gap to single particle excited states in the well is $\Delta_{1} \approx 0.24 \text{K}$, preserving the slope of $-2/\ln2$ throughout the accessible experimental range $20\text{mK}\lesssim T \lesssim 80\text{mK}$.
The $1$-channel ground state has lower energy than the $\psi$-channel by $ E_{1\psi} \approx 29\text{mK} $ in the absence of $V_X$, and the intrachannel gap  $\Delta^{1}_{\text{R}} \approx 220\text{mK}$, above the accessible range.
As $r_0 / l^*_B \approx 2.3$, we expect the contribution of $V_X$ to enhance the even-odd effect. 
Since the calculated charging energy is larger than the energy gap for the \nufh{} plateau, it is probably impossible to observe multiple transitions in a single well simply by changing the voltage on a back gate.
However, applying a voltage to a point-like gate on top of the sample can change the depth of a well by a large amount without inducing QPs in the surrounding $5/2$ state.
To further enhance the even-odd effect, all energy gaps need to be increased.
Increasing the charge on a point-gate or reducing the setback distance $d$ makes the confining trap tighter.
Increasing the magnetic length -- by lowering $B$ while maintaining the filling fraction -- increases all of the relevant splittings in a fixed trap geometry.

%
%
\paragraph{Conclusion}
The detection of non-Abelian QPs through local charge-sensing measurements falls within realistic experimental parameters. 
A sensitive compressibility measurement could extract slopes reflecting the degeneracies of the ground state.
Additional control over confinement potentials will allow for even more conclusive experiments.

%
%
\paragraph{Acknowledgments}
We thank B. Feldman, G. Gervais, S. Hart, V. Lahtinen, A. Stern, and V. Ventakachalam for useful discussions. G.B. supported by NSERC and FQRNT. C.R.L. supported by a Lawrence Gollub fellowship and the NSF through a grant for ITAMP at Harvard University.  I.N. supported by the Tel Aviv University Center for Nanoscience and Nanotechnology. This work was supported in part by a grant from the Microsoft Corporation, and by NSF grants DMR-0906475 and DMR-1206016.
%
%

\bibliography{majoranawellARXIV}
%
%
\appendix

%
%
%
%
\section{Appendix}

\section{Equilibration}
In the text, we estimate the equilibration time due to thermal excitation to be $t_{\text{T}} \sim 10^{-4}\text{s} \ll t_{\text{exp}}$. 
To obtain this estimate, we take $ t_{T} =  \hbar / E_{\text{T}} ,$
where $E_{\text{T}}\sim \Delta_{5/2}\exp \left( -\Delta_{5/2}/k_{B}T \right)$ is an Arrhenius estimate of thermally-induced inter-well hopping. 
Taking an activation gap of $\approx 0.25 \text{K}$ and a temperature of  $20 \text{mK}$, we obtain $t_T \sim 10^{-4} \text{s}$.
$t_{\text{exp}}$ is an experimental parameter, typically of order 10~Hz in existing measurements~\cite{Venkatachalam2011}.

\section{Matrix Elements}
The normalisation constant is: 
\begin{align*}
	\left\langle \ell | \ell \right\rangle &= \frac{1}{N_{\ell}}\int_0^{\infty}rdr \int_0^{\pi}d\phi \left( \frac{r}{l_B} \right)^{2l}e^{-r^2/4l_B^2}\overset{!}{=}1 \\
	\Rightarrow N_{\ell} &= \pi 2^{\ell} l_B^2 \ell !	
\end{align*}
The basis states are proportional to $e^{-r^2/8l_B^2}$ as opposed to the usual $e^{-r^2/4l_B^2}$ because $r$ is the relative coordinate.
The potential has diagonal terms, as well as an off-diagonal term only when circular symmetry is broken:
\begin{align}
	\bra{\ell} V^{REL}_{\text{trap}} \ket{\ell} &= \frac{(1+\alpha)}{2}k{l_B^*}^2(\ell+1) \nonumber\\
	\bra{\ell} V^{REL}_{\text{trap}} \ket{\ell + 2} &= \frac{(1-\alpha)}{4}{l_B^*}^2k\sqrt{(\ell+2)(\ell+1)} \nonumber\\
	\bra{\ell} V_{I} \ket{\ell} &=  \frac{{e^*}^2}{4\pi\epsilon {l_B^*} 2 } \frac{\Gamma [ \ell + 1/2 ] }{\Gamma [\ell+1]}.
\end{align}

\section{Saddle Point}
We demonstrate how to obtain the tunneling rate through a saddle point stated in the paper.
In the REL coordinate for two particles, the potential is:
\begin{equation}
	V(r)=\frac{\zeta}{r} + \frac{1}{2} \left( \frac{k}{2} \right) \left(x^2+\alpha y^2\right),
\end{equation}
with $\zeta={e^*}^2/4\pi\epsilon$ and effective magnetic length $\sqrt{2}l_B^*$.
In the circular case, $\alpha=1$, the minimum is circularly symmetric at $r_0=(2\zeta/k)^{1/3}$. 
For $\alpha \neq 1$, in the x-y plane, $(r_0,0)$ is still a minimum, and $(0,r_s)$ is a saddle point, with $r_s=r_0/(\alpha^{1/3})$.
Expanding to quadratic order near the minimum and the saddle point gives:
\begin{align}
	V(x-r_0,y)&=V(r_0)+\frac{1}{2} \frac{3k}{2} \left[ \delta x^2 + \left( \frac{\alpha}{3}-\frac{1}{3}\right)\delta y^2 \right], \nonumber \\
	V(0,y-r_s)&=V(r_s)+\frac{1}{2} \frac{3k\alpha}{2} \left[\left( \frac{1}{3\alpha}-\frac{1}{3}\right) \delta x^2 + \delta y^2 \right],
\end{align}
where as expected the transverse components vanish in the circular case, and the energy levels near $r_0$ are $E_n=V(r_0) + 3k {l_B^*}^2\sqrt{(\alpha-1)/3}(n+1/2)$.
We compare the ground state near $r_0$ to the height of the saddle point, $V(r_s)$, namely:
\begin{equation}
	\delta E = V(r_s)-E_0 = \frac{3k}{2}\left[ \frac{1}{2} r_0^2(\alpha^{1/3}-1 ) -{l_B^*}^2\sqrt{\frac{\alpha -1}{3}}\right].
\end{equation}
Solving for $\delta E > 0$ gives the condition $(r_0/l_B^*)^2 > \lambda(\alpha)$ described in the text.
To find the tunneling rate through the saddle point, we use the expression for the transmission through a saddle point potential $V_{SP}(x,y)=-U_xx^2+U_yy^2$ given in~\cite{FertigHalperin}, $T=(1+\exp(-\pi E))^{-1}$, where $E= -\delta E / \epsilon_1$, $\epsilon_1 = \sqrt{U_x U_y}/m\omega_c$ in the large $B$-limit, and $\omega_c$ is the cyclotron frequency. 
For the saddle point under consideration, $m\omega_c = 1/2{l_B^*}^2$, and $\epsilon_1 = \frac{k}{2}\sqrt{3\alpha(\alpha-1)}{l_B^*}^2$, giving:
\begin{equation}
	E = \frac{-3\left[ \frac{1}{2}r_0^2(\alpha^{1/3}-1)-{l_B^*}^2\sqrt{\frac{\alpha - 1}{3}} \right]}{\sqrt{3 \alpha (\alpha-1)}{l_B^*}^2},
\end{equation}
which reduces to $E \sim \frac{-1}{2\sqrt{3\alpha}}\left( \frac{r_0}{l_B^*} \right)^2\frac{\alpha^{1/3}-1}{\sqrt{\alpha-1}}$ for large $r_0/l_B^*$, and in this limit, $T \sim \exp(\pi E)$.
To convert from transmission probability to a transmission rate, we take the velocity of a QP about its orbit as the ratio of the gradient of the potential to the magnetic field, and dividing by the circumference of an orbit near the minimum, we find the frequency of the orbit is $\sim k{l_B^*}^2$, which, multiplied by $T$, gives the rate $R$ reported in the main text.

\section{2-particle Splittings}
\subsection{Maximum Values}
The maximum values for the  gaps $\Delta^f$ and $E_{1\psi}$ in the two-particle REL spectrum are given in the text. The pre-factors are found by expanding the potential near the minimum at $r_0$, finding the lowest-energy and first excited states in terms of allowed angular momenta, and expanding terms as a function of $l_B^*/r_0 \ll 1$. 
To find the angular momentum corresponding to the lowest energy state, we use the relation $r=\sqrt{2\ell} l_B^*$ to find $\ell_0$, the (possibly not allowed) angular momentum corresponding to $r_0$, we first find the allowed angular momenta right above and below: 
\begin{align}
	\ell_{-}&=2\lfloor(\ell-\theta_f/\pi )/2\rfloor + \theta_f/\pi  \\
	\ell_{+}&=2(\lfloor(\ell-\theta_f/\pi )/2\rfloor +1) + \theta_f/\pi,
\end{align}
where $\lfloor \cdot \rfloor$ is the integer floor function.
Next, convert back to positions corresponding to these momenta, and plug back into the potential to check which state has lower energy. 
If $\ell_g$ is the angular momentum of the ground state, then the angular momentum of the first excited state is $\ell_1 = \ell_g \pm 2$. 
This argument is sufficient for crudely extracting the large $r_0$ behaviour of the intra-channel gap $\Delta^f$.
A similar calculation produces $E_{1\psi}$.
\begin{figure}[htbp]
	\centering
		\includegraphics[scale=1]{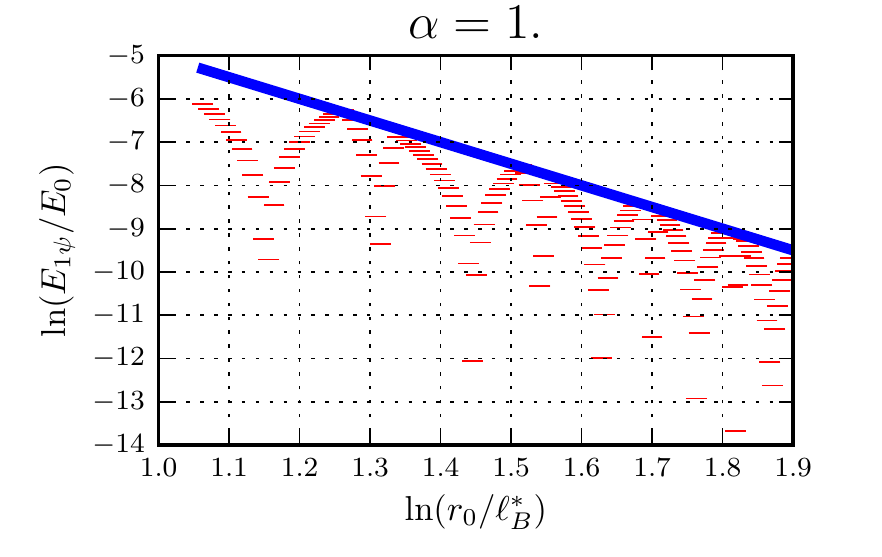}
	\caption{$ \log(E_{1\psi} / E_0) $ calculated numerically by exact diagonalisation, vs. $ \log(r_0/l_B^* )$ for a circular well. A power law $r_0^{-5}$ is also plotted as a guide to the eye. $E_0 = \frac{9}{2} \frac{{e^*}^2}{4\pi \epsilon} \frac{1}{l_B^*} $, as described in the paper below equation 8, with the $r_0 / l_B^*$ dependence factored out.}
	\label{fig:CircularSplitting}
\end{figure}
\begin{figure}[htbp]
	\centering
		\includegraphics[scale=1]{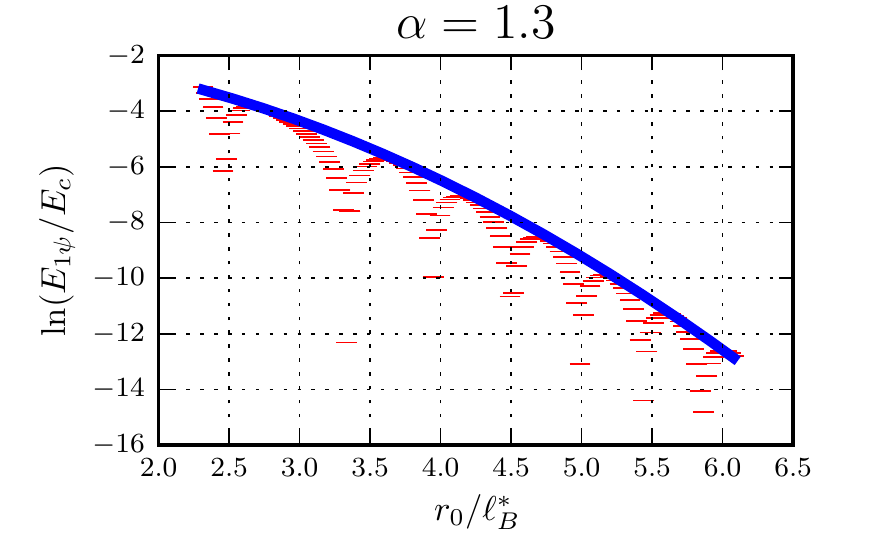}
	\caption{$\log(E_{1\psi}/E_c)$ calculated numerically by exact diagonalisation, vs. $r_0/l_B^*$ for an elliptical well, with $\alpha=1.3$. A Gaussian decay $\exp\left(-c r_0^2  \right)$ is also plotted as a guide to the eye. $E_c =  \frac{{e^*}^2}{4\pi \epsilon} \frac{1}{l_B^*}$ is the Coulomb energy scale. }
	\label{fig:EllipticalSplitting}
\end{figure}
The decay of $E_{1\psi}$ is plotted in figure~\ref{fig:CircularSplitting} for a circular well; the log-log plot demonstrates a power law decay as $r_0^{-5}$.  
Figure~\ref{fig:EllipticalSplitting} shows the decay of $E_{1\psi}$ for an elliptical well with $\alpha = 1.3$; the quadratic decay in a log plot confirms a Gaussian form as expected from the saddle-point calculation.

\section{Relations between $\kappa$ , the local incompressibility, and the  SET signal. }

\subsection{Chemical Potential v. Backgate}
In this letter, we calculate $\kappa$, the change in local average number of QPs as a function of the local chemical potential of the QPs. 
Charge stability diagrams are then drawn as a function of the QP chemical potential. 
The proposed SET measurements reveal the \textit{inverse compressibility} by measuring the change of the local electrostatic potential in the 2DEG as a function of a global backgate voltage. 
The electrostatic potential is then converted to chemical potential of electrons, under the assumption that electrochemical potential is held constant. 
A capacitance model is used to convert backgate voltage to average electron density.
The qualitative even-odd effect is independent of definitions, but the quantitative slope predicted needs to be scaled to match experimental parameters.

\subsection*{Relation of the SET signal to the local incompressibility}

We elaborate here on the  relation between the local quasiparticle compressibility, discussed  in this paper, and the signal measured in an SET experiment, which is commonly described as measuring the local electronic {\em incompressibility}. 
Throughout the remainder of the appendix, we consider a sample with its top surface at $z=0$. A 2DEG is found a distance $d$ below the surface, and a backgate is placed a distance $D$ below the 2DEG, at $z=-(d+D)$. The SET will be placed just above the sample surface, $z = 0^+$.

A more precise description of the SET signal is that it measures 
 $\partial \Phi_{\rm{SET}} / \partial V_G $, where  $ \Phi_{\rm{SET}} $ is the electrostatic potential at the SET,   and    $V_G$ is the back-gate voltage. In this measurement,   the electrochemical potential of  the 2DEG  is held fixed at a voltage $V$, by connecting it to a metallic lead.  (In ref~\cite{Venkatachalam2011}, the lead is grounded; i.e., the 2DEG is in equilibrium with a ground surface at infinity, and we may take $V=0$.)   Furthermore, 
$$
\delta   \Phi_{\rm{SET}} = \int d^2 \vec{r} \, K(\vec{r} - \vec{r}_0)\, \delta \Phi (\vec{r}),
$$
where $ \Phi({\vec{r}})$ is the electrostatic  potential at a point $\vec{r}$ just above the plane of the 2DEG,  inside the GaAs, $\vec{r}_0$ denotes the horizontal location of the SET probe,  and the precise form of the kernel $K$ depends on the height of the SET probe above the semiconductor surface, the depth of the 2DEG, and dielectric constant $\epsilon$ of the material. In general, the fluctuation in $ \Phi_{\rm{SET}}$ may be interpreted as a weighted spatial average of $\Phi$ within a distance of the order of the SET-2DEG separation, $d$.   

It is customary to define a local chemical potential for electrons in the 2DEG, by 
$$
\mu_e(\vec{r}) \equiv e [ V - \Phi(\vec{r}) ] ,
$$
where $e<0$ is the electron charge.  
The change in the average density of electrons in the 2DEG produced by a change in the back-gate voltage is
$$
 \delta \bar{n} =  - C \delta V_G /e,
$$
where C is an effective capacitance per unit area.
 Then, if we define the local  electronic incompressibility by $\gamma(\vec{r}) \equiv\partial \mu_e (\vec{r}) / \partial \bar{n}$, we see that 
$$
\frac {\partial   \Phi_{\rm{SET}}}{\partial V_G} = \frac{C}{e^2}  \int d^2 \vec{r} \, K(\vec{r} - \vec{r}_0)\, \gamma  (\vec{r}),
$$
Moreover, if we define $\bar{\gamma}$ as the spatial average of the local incompressibility $\gamma (\vec{r})$, one finds 
$$
C^{-1} = \frac{D}{\epsilon} + \frac {\bar{\gamma}}{e^2},
$$
where $D$ is the distance to the back gate.
We note that $\bar{\gamma}$ will be finite, even when the bulk of the system sits in the quantum Hall plateau, due to the effects of changing  quasiparticle populations in wells whose depths are close to a critical value.  In practice, in the experimental geometry where $D$ is the order of a micrometer, the   first term will be large compared to the second, and $C$ will be determined primarily by the geometric capacitance.

\subsection*{Relation between the SET signal and the local quasiparticle compressibility}

As a consequence of Poisson's equation, the value of $\Phi(\vec{r})$ will be directly affected by changes in the local electron density $n(\vec{r})$ .  In the simplest case, we consider a situation where there is a single chargeable potential well,  surrounded by a region of incompressible 5/2 state, in the area sensed by the SET.  Then changes in the local electron density result primarily from changes in $N$, the quasiparticle occupation number of the well.  To a good approximation, 
$$
\frac {\partial   \Phi_{\rm{SET}}}{\partial V_G} \approx \eta(d) \frac{\delta N}{\delta V_G}  + c_1,
$$
where $\eta$ is the model-dependent potential at the SET from a single QP, and $c_1$  is a slowly varying number accounting for the change in density in the rest of the sample beyond the well.  

The quasiparticle chemical potential $\mu$  employed in our paper is related in a complicated way to the local electron chemical potential $\mu_e$.  Roughly,  
$$
\delta \mu =  - e^* \delta \tilde {\Phi}
$$
where $\tilde {\Phi} $ is the electrostatic potential at the position of the well, {\em excluding any potential due to the presence of one or more quasiparticles in the well.}  When the expectation value of $N$ is a rapidly varying quantity, there will  be a  very large difference between the variation in $\tilde {\Phi} $ and the variation in 
$\Phi(\vec{r})$. 

In general, we expect that $\tilde {\Phi} $  should depend smoothly on the back-gate voltage $V_G$.  We may estimate this dependence by assuming that the potential well is surrounded by an incompressible region of radius $R$, and that outside this region we have a continuous medium characterized by a finite incompressibility $\bar {\gamma}$, which we identify with the spatial average of $\gamma(\vec{r})$ defined above.  We shall assume that $D$ is very large compared to $R$ and to the ``screening length"
$l_s \equiv \bar{\gamma} \epsilon / e^2$, but we should still consider different values of the ratio $R/l_s$.  
In the case $R > l_s$, analysis of the resulting electrostatics problem leads to a result 
$$
\delta \tilde{\Phi} \sim c_2 \delta V_G \,  R/ D 
$$
 where $c_2$ is a constant of order unity.
 In the opposite limit, $l_s > R$, we find 
 $$
 \delta \tilde{\Phi} \sim  \delta V_g l_s/D =  \delta V_G \bar {\gamma} / Ce^2 .
 $$
When the system is in the middle of a quantum Hall plateau, we expect that $\bar{\gamma}$ will be large, and we might expect to  be in the regime $l_s>R$.

 Finally we may put these results together to find the relation between the SET signal and the quasiparticle compressibility defined in the text.  In the limit where  $l_s > R$,  we have, ignoring a smooth background contribution, 
 $$
	\frac {\partial   \Phi_{\rm{SET}}}{\partial V_G} \approx -  \left( \frac {e^*} {e} \right)^2
 	\frac{\epsilon \, \bar{\gamma} \, \kappa}{D} \frac{\eta(d)}{e^*} .
 $$
 
 The above  model can be generalized to a situation where there are several wells beneath the SET tip, by choosing a larger radius R within which there is no continuum background compressibility, and including explicitly  the Coulomb interactions between quasiparticles in different wells in this region. In the simpler model we have replaced all wells by a continuum, except for the one under consideration.
 
\subsection*{Calculation of $\eta(d)$}
We calculate $\eta(d)$ for a point-like SET, located an infinitesimal distance above the sample surface, which is a distance $d$ from the 2DEG.
The distance to the backgate, $D$ is taken to be much larger than $d$.
We consider a QP of charge $e^*$ added to the 2DEG in a bulk sample with permittivity $\epsilon$, and we want to know the potential at the SET, when the sample sits in vacuum - i.e outside, permittivity is $\epsilon_0$.
We have to solve the following equations for an electric field $\vec{E}$:
\begin{align*}
	\epsilon \vec{\nabla} \cdot \vec{E} &= \rho,~~~z<0 \\
	\epsilon_0 \vec{\nabla} \cdot \vec{E} &= 0, ~~~z>0 \\
	\vec{\nabla} \times \vec{E} &= 0, ~~~\text{everywhere},
\end{align*}
with boundary conditions at the sample boundary (i.e. $z=0$) of continuous $\vec{E}$-fields in the $x$ and $y$ directions, and $ \displaystyle \lim_{z \rightarrow 0^+} \epsilon_0 E_z = \lim_{z \rightarrow 0^-} \epsilon E_z$.
Place an image charge $q'$ at $z=d$ above the surface, and then using cylindrical coordinates $(r,\phi, z)$, the potential at any point inside the sample is:
\begin{equation}
	\phi = \frac{1}{\pi \epsilon}\left( \frac{e^*}{R_1} + \frac{q'}{R_2}\right), ~~~~ z<0,
\label{eq:SuppPhiIN}
\end{equation}
where $R_1=\sqrt{r^2+(d+z)^2}$, $R_2=\sqrt{r^2+(d-z)^2}$.
For the region $z>0$, which is where the SET is, there are no charges, and the potential must therefore be a solution to Laplace's equation without singularities. 
The simplest solution is the potential from an effective charge $q$ located at the site of the QP $e^*$, giving a potential:
\begin{equation}
	\phi = \frac{1}{4\pi \epsilon_0} \frac{q}{R_1}, ~~~ z>0.
\label{eq:SuppPhiOUT}
\end{equation}
The solutions~\ref{eq:SuppPhiIN} and~\ref{eq:SuppPhiOUT} can be matched at $z=0$ and must satisfy the boundary conditions, giving  $e^*-q'=q$ and $(e^*+q')/\epsilon=q/\epsilon_0$.
This implies $q=\left(2\epsilon_0/\epsilon+\epsilon_0  \right) e^*$.
We thus find the potential at the SET due to the QP to be:
$$
	\eta(d)= \frac{1}{4\pi}\frac{2}{\epsilon+\epsilon_0}\frac{e^*}{d}.
$$
A very similar calculation can be found in section 4.4 in~\cite{Jackson}.


%
\end{document}